\documentclass[aps,prd,twocolumn,showpacs,groupedaddress]{revtex4}  
\usepackage{graphicx}  
\usepackage{dcolumn}   
\usepackage{bm}        
\usepackage{amssymb}   
\usepackage{axodraw}

\begin{document}


\hspace{5.2in} \mbox{Fermilab-Pub-06/081-E}

\title{Search for Excited Muons in $p\bar{p}$ Collisions at $\sqrt{s} = 1.96$~TeV}
%
\author{                                                                      
V.M.~Abazov,$^{36}$                                                           
B.~Abbott,$^{76}$                                                             
M.~Abolins,$^{66}$                                                            
B.S.~Acharya,$^{29}$                                                          
M.~Adams,$^{52}$                                                              
T.~Adams,$^{50}$                                                              
M.~Agelou,$^{18}$                                                             
J.-L.~Agram,$^{19}$                                                           
S.H.~Ahn,$^{31}$                                                              
M.~Ahsan,$^{60}$
% ...
\\                                                                            
\vskip 0.30cm                                                                 
\centerline{(D\O\ Collaboration)}                                             
\vskip 0.30cm                                                                 
}                                                                             
\affiliation{                                                                 
\centerline{$^{1}$Universidad de Buenos Aires, Buenos Aires, Argentina}       
\centerline{$^{2}$LAFEX, Centro Brasileiro de Pesquisas F{\'\i}sicas,         
                  Rio de Janeiro, Brazil}                                     
\centerline{$^{3}$Universidade do Estado do Rio de Janeiro,                   
                  Rio de Janeiro, Brazil}                                     
\centerline{$^{4}$Instituto de F\'{\i}sica Te\'orica, Universidade            
                  Estadual Paulista, S\~ao Paulo, Brazil}                     
\centerline{$^{5}$University of Alberta, Edmonton, Alberta, Canada,           
                  Simon Fraser University, Burnaby, British Columbia, Canada,}
\centerline{York University, Toronto, Ontario, Canada, and                    
                  McGill University, Montreal, Quebec, Canada}                
\centerline{$^{6}$Institute of High Energy Physics, Beijing,                  
                  People's Republic of China}                                 
\centerline{$^{7}$University of Science and Technology of China, Hefei,       
                  People's Republic of China}                                 
\centerline{$^{8}$Universidad de los Andes, Bogot\'{a}, Colombia}             
\centerline{$^{9}$Center for Particle Physics, Charles University,            
                  Prague, Czech Republic}                                     
\centerline{$^{10}$Czech Technical University, Prague, Czech Republic}        
\centerline{$^{11}$Center for Particle Physics, Institute of Physics,         
                   Academy of Sciences of the Czech Republic,                 
                   Prague, Czech Republic}                                    
\centerline{$^{12}$Universidad San Francisco de Quito, Quito, Ecuador}        
\centerline{$^{13}$Laboratoire de Physique Corpusculaire, IN2P3-CNRS,         
                   Universit\'e Blaise Pascal, Clermont-Ferrand, France}      
\centerline{$^{14}$Laboratoire de Physique Subatomique et de Cosmologie,      
                   IN2P3-CNRS, Universite de Grenoble 1, Grenoble, France}    
\centerline{$^{15}$CPPM, IN2P3-CNRS, Universit\'e de la M\'editerran\'ee,     
                   Marseille, France}                                         
\centerline{$^{16}$IN2P3-CNRS, Laboratoire de l'Acc\'el\'erateur              
                   Lin\'eaire, Orsay, France}                                 
\centerline{$^{17}$LPNHE, IN2P3-CNRS, Universit\'es Paris VI and VII,         
                   Paris, France}                                             
\centerline{$^{18}$DAPNIA/Service de Physique des Particules, CEA, Saclay,    
                   France}                                                    
\centerline{$^{19}$IReS, IN2P3-CNRS, Universit\'e Louis Pasteur, Strasbourg,  
                    France, and Universit\'e de Haute Alsace,                 
                    Mulhouse, France}                                         
\centerline{$^{20}$Institut de Physique Nucl\'eaire de Lyon, IN2P3-CNRS,      
                   Universit\'e Claude Bernard, Villeurbanne, France}         
\centerline{$^{21}$III. Physikalisches Institut A, RWTH Aachen,               
                   Aachen, Germany}                                           
\centerline{$^{22}$Physikalisches Institut, Universit{\"a}t Bonn,             
                   Bonn, Germany}                                             
\centerline{$^{23}$Physikalisches Institut, Universit{\"a}t Freiburg,         
                   Freiburg, Germany}                                         
\centerline{$^{24}$Institut f{\"u}r Physik, Universit{\"a}t Mainz,            
                   Mainz, Germany}                                            
\centerline{$^{25}$Ludwig-Maximilians-Universit{\"a}t M{\"u}nchen,            
                   M{\"u}nchen, Germany}                                      
\centerline{$^{26}$Fachbereich Physik, University of Wuppertal,               
                   Wuppertal, Germany}                                        
\centerline{$^{27}$Panjab University, Chandigarh, India}                      
\centerline{$^{28}$Delhi University, Delhi, India}                            
\centerline{$^{29}$Tata Institute of Fundamental Research, Mumbai, India}     
\centerline{$^{30}$University College Dublin, Dublin, Ireland}                
\centerline{$^{31}$Korea Detector Laboratory, Korea University,               
                   Seoul, Korea}                                              
\centerline{$^{32}$SungKyunKwan University, Suwon, Korea}                     
\centerline{$^{33}$CINVESTAV, Mexico City, Mexico}                            
\centerline{$^{34}$FOM-Institute NIKHEF and University of                     
                   Amsterdam/NIKHEF, Amsterdam, The Netherlands}              
\centerline{$^{35}$Radboud University Nijmegen/NIKHEF, Nijmegen, The          
                  Netherlands}                                                
\centerline{$^{36}$Joint Institute for Nuclear Research, Dubna, Russia}       
\centerline{$^{37}$Institute for Theoretical and Experimental Physics,        
                   Moscow, Russia}                                            
\centerline{$^{38}$Moscow State University, Moscow, Russia}                   
\centerline{$^{39}$Institute for High Energy Physics, Protvino, Russia}       
\centerline{$^{40}$Petersburg Nuclear Physics Institute,                      
                   St. Petersburg, Russia}                                    
\centerline{$^{41}$Lund University, Lund, Sweden, Royal Institute of          
                   Technology and Stockholm University, Stockholm,            
                   Sweden, and}                                               
\centerline{Uppsala University, Uppsala, Sweden}                              
\centerline{$^{42}$Physik Institut der Universit{\"a}t Z{\"u}rich,            
                   Z{\"u}rich, Switzerland}                                   
\centerline{$^{43}$Lancaster University, Lancaster, United Kingdom}           
\centerline{$^{44}$Imperial College, London, United Kingdom}                  
\centerline{$^{45}$University of Manchester, Manchester, United Kingdom}      
\centerline{$^{46}$University of Arizona, Tucson, Arizona 85721, USA}         
\centerline{$^{47}$Lawrence Berkeley National Laboratory and University of    
                   California, Berkeley, California 94720, USA}               
\centerline{$^{48}$California State University, Fresno, California 93740, USA}
\centerline{$^{49}$University of California, Riverside, California 92521, USA}
\centerline{$^{50}$Florida State University, Tallahassee, Florida 32306, USA} 
\centerline{$^{51}$Fermi National Accelerator Laboratory,                     
            Batavia, Illinois 60510, USA}                                     
\centerline{$^{52}$University of Illinois at Chicago,                         
            Chicago, Illinois 60607, USA}                                     
\centerline{$^{53}$Northern Illinois University, DeKalb, Illinois 60115, USA} 
\centerline{$^{54}$Northwestern University, Evanston, Illinois 60208, USA}    
\centerline{$^{55}$Indiana University, Bloomington, Indiana 47405, USA}       
\centerline{$^{56}$University of Notre Dame, Notre Dame, Indiana 46556, USA}  
\centerline{$^{57}$Purdue University Calumet, Hammond, Indiana 46323, USA}    
\centerline{$^{58}$Iowa State University, Ames, Iowa 50011, USA}              
\centerline{$^{59}$University of Kansas, Lawrence, Kansas 66045, USA}         
\centerline{$^{60}$Kansas State University, Manhattan, Kansas 66506, USA}     
\centerline{$^{61}$Louisiana Tech University, Ruston, Louisiana 71272, USA}   
\centerline{$^{62}$University of Maryland, College Park, Maryland 20742, USA} 
\centerline{$^{63}$Boston University, Boston, Massachusetts 02215, USA}       
\centerline{$^{64}$Northeastern University, Boston, Massachusetts 02115, USA} 
\centerline{$^{65}$University of Michigan, Ann Arbor, Michigan 48109, USA}    
\centerline{$^{66}$Michigan State University,                                 
            East Lansing, Michigan 48824, USA}                                
\centerline{$^{67}$University of Mississippi,                                 
            University, Mississippi 38677, USA}                               
\centerline{$^{68}$University of Nebraska, Lincoln, Nebraska 68588, USA}      
\centerline{$^{69}$Princeton University, Princeton, New Jersey 08544, USA}    
\centerline{$^{70}$State University of New York, Buffalo, New York 14260, USA}
\centerline{$^{71}$Columbia University, New York, New York 10027, USA}        
\centerline{$^{72}$University of Rochester, Rochester, New York 14627, USA}   
\centerline{$^{73}$State University of New York,                              
            Stony Brook, New York 11794, USA}                                 
\centerline{$^{74}$Brookhaven National Laboratory, Upton, New York 11973, USA}
\centerline{$^{75}$Langston University, Langston, Oklahoma 73050, USA}        
\centerline{$^{76}$University of Oklahoma, Norman, Oklahoma 73019, USA}       
\centerline{$^{77}$Oklahoma State University, Stillwater, Oklahoma 74078, USA}
\centerline{$^{78}$Brown University, Providence, Rhode Island 02912, USA}     
\centerline{$^{79}$University of Texas, Arlington, Texas 76019, USA}          
\centerline{$^{80}$Southern Methodist University, Dallas, Texas 75275, USA}   
\centerline{$^{81}$Rice University, Houston, Texas 77005, USA}                
\centerline{$^{82}$University of Virginia, Charlottesville,                   
            Virginia 22901, USA}                                              
\centerline{$^{83}$University of Washington, Seattle, Washington 98195, USA}  
}                                                                             
\date{April 19, 2006}

\begin{abstract}
We present the results of a search for the production of an excited state of the muon,
$\mu^*$, in proton antiproton collisions at $\sqrt s = 1.96$~TeV. The data have been
collected with the D0 experiment at the Fermilab Tevatron Collider and correspond to an
integrated luminosity of approximately $380$~pb$^{-1}$. We search  for $\mu^*$ in the process
$p \bar p \rightarrow \mu^* \mu$, with the $\mu^*$ subsequently decaying to a muon plus
photon. No excess above the standard model expectation is observed in data. Interpreting
our data in the context of a model that describes $\mu^*$ production by four-fermion
contact interactions and $\mu^*$ decay via electroweak processes, we exclude production
cross sections higher than $0.057$~pb -- $0.112$~pb at the 95\% confidence level, depending on
the mass of the excited muon. Choosing the scale for contact interactions to be
$\Lambda = 1$~TeV, excited muon masses below 618~GeV are excluded.
\end{abstract}

\maketitle

An open question in particle physics is the observed mass hierarchy of the
quark and lepton SU(2) doublets in the standard model (SM). A commonly
proposed explanation for the three generations is a compositeness model
\cite{compositeness} of the known leptons and quarks. According to this
approach, a quark or lepton is a bound state of three fermions, or of a
fermion and a boson \cite{Terazawa1}.
Due to the underlying substructure, compositeness models imply a large spectrum
of excited states. 
The coupling of excited fermions to ordinary quarks and leptons, resulting
from novel strong interactions, can be described by contact interactions (CI) with
the effective four-fermion Lagrangian \cite{Baur90}
$$
 \mathcal{L}_{\mathrm{CI}} \, = \, \frac{g^2}{2\Lambda^2}\, j^{\mu}\, j_{\mu},
$$
where $j_{\mu}$ is the fermion current
\begin{eqnarray*}
  j_{\mu} & = &  \eta_L \, \bar{f}_L\gamma_{\mu}f_L + \eta'_L \, \bar{f}^*_L\gamma_{\mu}f^*_L + \eta''_L \, \bar{f}^*_L \gamma_{\mu}f_L \\
  & & + \, h.c. \, + \, (L \rightarrow R).
\end{eqnarray*}
The SM and excited fermions are denoted by $f$ and $f^*$, respectively; $g^2$ is
chosen to be $4 \pi$, the $\eta$ factors for the left-handed currents are
conventionally set to one, and the right-handed currents are set to zero. The
compositeness scale is $\Lambda$.

Gauge mediated transitions between ordinary and excited fermions can be described by the
effective Lagrangian \cite{Baur90}
\begin{eqnarray*}
  \mathcal{L}_{\mathrm{EW}} & & = \ \frac{1}{2\Lambda} \bar{f}_R^*\,\sigma^{\mu\nu} \\
     & & \left[ g_s f_s \frac{\lambda^a}{2} G^a_{\mu\nu} + g f \frac{\tau}{2} W_{\mu\nu} 
                    +   g'f'\frac{Y}{2}B_{\mu\nu}\right] f_L + h.c.
\end{eqnarray*}
where $G^a_{\mu\nu}$, $W_{\mu\nu}$, and $B_{\mu\nu}$ are the field strength tensors
of the gluon, the SU(2) and U(1) gauge fields, respectively; $f_s$, $f$ and $f'$
are parameters of order one.



\begin{figure}[ht]
\begin{picture}(245,50)(8,0)
    \SetScale{0.40}
    \SetWidth{1.5}
    \SetColor{Black}
    \ArrowLine(25,10)(100,60)
    \ArrowLine(25,110)(100,60)
    \ArrowLine(100,60)(175,10)
    \ArrowLine(270,60)(345,10)
    \Text(136,37)[t]{$\gamma$}
    \Text(136,16)[t]{$\mu$}
    \Text(68,16)[t]{$\mu$}
    \Text(12,16)[t]{$q$}
    \Text(12,37)[t]{$\bar{q}$}
    \Text(103,39)[t]{$\mu^*$}
    \Text(68,37)[t]{$\mu^*$}
    \Photon(270,60)(345,110){3.8}{5.9}
    \SetWidth{2.5}
    \SetColor{Fuchsia}
    \ArrowLine(100,60)(175,110)
    \ArrowLine(195,75)(270,60)
    \SetColor{Red}
    \Vertex(100,60){3}
    \Vertex(270,60){3}
    \put(150.0,-8.0) {\includegraphics[scale=0.33]{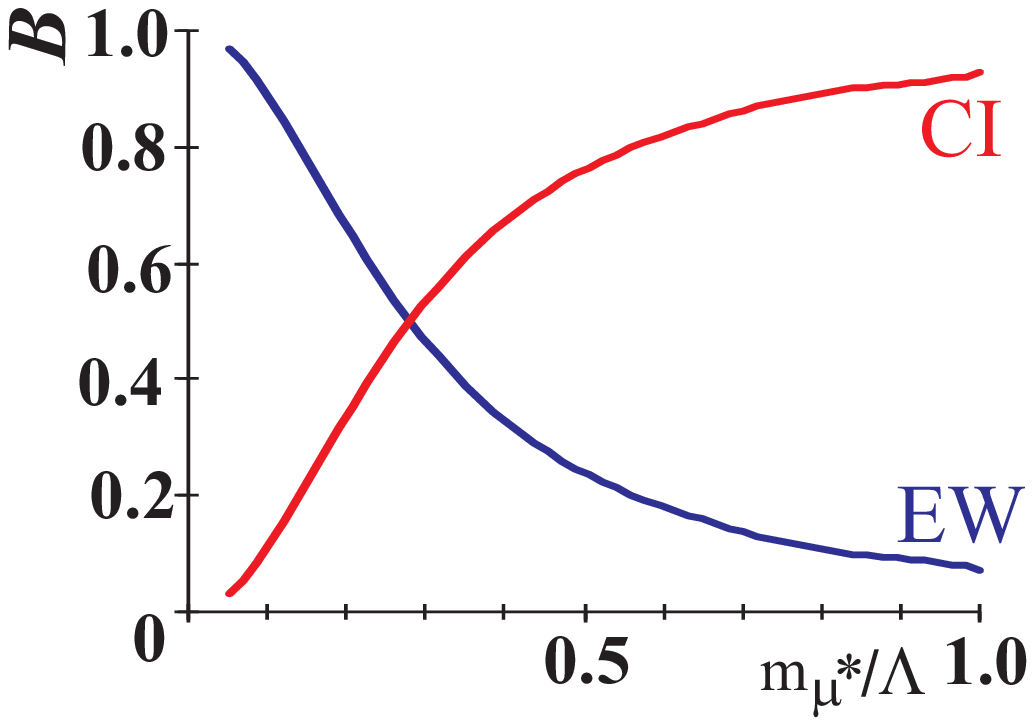}}
    \end{picture}
  \caption{Four-fermion contact interaction $q \bar{q} \rightarrow \mu^* \mu$, and electroweak decay
           $\mu^* \rightarrow \mu \gamma$. On the right, the relative contribution of decays via
	   CI and via electroweak interactions (EW) as a function of $m_{\mu^*} / \Lambda$ is shown.}
  \label{feynman}
\end{figure}

The present analysis considers single production of an excited muon $\mu^*$ in association
with a muon via four-fermion CI, with the subsequent electroweak decay of the $\mu^*$ into a
muon and a photon (Fig.~\ref{feynman}). This decay mode leads to the fully reconstructable and
almost background-free final state $\mu \mu \gamma$.
With the data considered herein, collected with the D0 detector at the
Fermilab Tevatron Collider in $p\bar{p}$ collisions at $\sqrt s = 1.96$~TeV, the largest
expected SM background is from the Drell-Yan (DY) process
$p\bar{p}\rightarrow Z/\gamma^* \rightarrow \mu^+\mu^- (\gamma)$, with the final state
photon radiated by either a parton in the initial state $p$ or $\bar p$, or from one
of the final state muons. This background can be strongly suppressed by the application
of suitable selection criteria. Other backgrounds are small.

Excited muons have been searched for unsuccessfully previously \cite{pdg}, e.g.~at the LEP
$e^+e^-$ collider; however the reach has been limited by the center-of-mass energy available
to $m_{\mu^*} < 190$~GeV. Searches for quark-lepton compositeness via deviations from the
Drell-Yan cross section have excluded values of $\Lambda$ of up to $\approx 6$~TeV depending
on the chirality \cite{comprun1}. The present analysis is complementary to those results in
the sense that an exclusive channel and different couplings ($\eta$ factors) are probed. The
CDF collaboration has recently presented results \cite{cdf} for the production of excited
electrons which will be discussed later.

For the simulation of the signal a customized version of the {\sc pythia} event generator
\cite{pythia} is used, following the model of \cite{Baur90}. The branching fraction for
the decay $\mu^* \rightarrow \mu \gamma$ normalized to all gauge particle decay modes is
30\% for masses above 300 GeV, and for smaller $\mu^*$ masses it increases up to 73\% at
$m_{\mu^*} = 100$~GeV. Decays via contact interactions, not implemented in {\sc pythia},
contribute between a few percent of all decays for $\Lambda \gg m_{\mu^*}$ and 92\% for
$\Lambda = m_{\mu^*}$ \cite{Baur90,LHC} (see Fig.~\ref{feynman}). This has been taken
into account for the signal expectation. The leading order cross section calculated with
{\sc pythia} has been corrected to next-to-next-to-leading order (NNLO) \cite{dy,daleo};
the corresponding correction factor varies between 1.430 (1.468) for $m_{\mu^*} = 100$~GeV (200~GeV)
and 1.312 for $m_{\mu^*} = 1$~TeV. The total width is greater than 1~GeV for
$100$~GeV~$\leq m_{\mu^*} \leq 1000$~GeV, thus lifetime effects can be neglected. For the
values of $m_{\mu^*}$ and $\Lambda$ studied here, the total width is always less than
10\% of $m_{\mu^*}$ \cite{Baur90}.

The dominant SM background process at all stages of the selection is DY production of $\mu^+\mu^-$
pairs. This background, as well as diboson ($WW$, $WZ$, $ZZ$) production, has been
simulated with the {\sc pythia} Monte Carlo (MC) program. The DY expectation has been
corrected using the NNLO calculation from \cite{dy}. For diboson production, the
next-to-leading order cross sections from \cite{mcfm} are used. Monte Carlo events, both
for SM and signal, have been passed through a detector simulation based on the
{\sc geant} \cite{geant} package,
and reconstructed using the same reconstruction program as the data.
The CTEQ5L parton distribution functions (PDF) \cite{cteq} are used for the generation
of all MC samples.

The analysis is based on the data collected with the D0 detector \cite{run2det} between
August 2002 and September 2004, corresponding to an integrated luminosity of
$380$~pb$^{-1}$. The D0 detector includes a central tracking
system, comprised of a silicon microstrip tracker (SMT) and a central fiber tracker
(CFT), both located within a 2~T superconducting solenoidal magnet. The SMT has
$\approx 800,000$ individual strips, with typical pitch of $50-80$ $\mu$m, and a design
optimized for tracking and vertexing capability at pseudorapidities \cite{coordinate}
of $|\eta|<2.5$. The CFT has eight coaxial barrels, each supporting two doublets of
scintillating fibers of 0.835~mm diameter, one doublet being parallel to the collision
axis, and the other alternating by $\pm 3^{\circ}$ relative to the axis. Three liquid
argon and uranium calorimeters provide coverage out to $|\eta|\approx 4.2$: a central
section covering $|\eta|$ up to $\approx 1.1$, and two end calorimeters. A muon system
resides beyond the calorimetry, and consists of a layer of tracking detectors and
scintillation trigger counters before 1.8~T iron toroids, followed by two similar layers
after the toroids. Tracking at $|\eta|<1$ relies on 10~cm wide drift tubes, while 1~cm
mini-drift tubes are used at $1<|\eta|<2$.
Luminosity is measured using 
scintillator arrays located in front of the end calorimeter cryostats, covering
$2.7 < |\eta| < 4.4$.

Trigger and data acquisition systems are designed to accommodate the high
luminosities of the Tevatron Run II. Based on information from tracking,
calorimetry, and muon systems, the output of the first two levels of the trigger
is used to limit the rate for accepted events to $< 1$~kHz,
relying on hardware and firmware. The third and final level
of the trigger uses software algorithms and a computing farm to reduce the output
rate to $\approx$ 50~Hz, which is written to tape.

Efficiencies for muon and photon identification and track reconstruction are
determined from the simulation. To verify the simulation and to estimate systematic
uncertainties, the efficiencies have also been calculated from data samples, using $Z
\rightarrow \mu^+\mu^-$ candidate events and inclusive dimuon events for muons
and tracks, and $Z \rightarrow e^+e^-$ events to determine the efficiency of
reconstructing electrons.
We assume that the different response for electrons and photons in the
calorimeter is properly modelled by the simulation. The transverse (with respect
to the beam axis) momentum resolution of the central tracker and the energy
resolution of the calorimeter have been tuned in the simulation to reproduce the
resolutions observed in the data using $Z\rightarrow \ell \ell$ ($\ell = e, \mu$)
events.

The process $p \bar p \rightarrow \mu^* \mu$ with $\mu^* \rightarrow \mu \gamma$
leads to a final state with two highly energetic isolated muons and a photon. We
require two muons to be identified in the muon system and each matched to a track in
the central tracking system with 
transverse momentum $p_T > 15$~GeV. The events have been collected with Level 1
trigger conditions requiring two muons detected by the muon scintillation counters,
with at least one muon with tightened criteria identified by the Level 2 trigger,
and requiring a segment reconstructed in the muon system above certain $p_T$ thresholds
and/or a track in the central tracking system above certain $p_T$ thresholds at Level 3.
The trigger efficiency has been determined from independent data samples for each
trigger object (muon) and trigger level separately. The overall trigger
efficiency which is applied to the simulation is found to be $88 \pm 6$\%
for the signal after application of all selection criteria.

\begin{figure} \setlength{\unitlength}{1cm}
  \begin{picture}(8.6,4.2)(0.0,0.0)
  \put(-0.3,-0.3) {\includegraphics[scale=0.23]{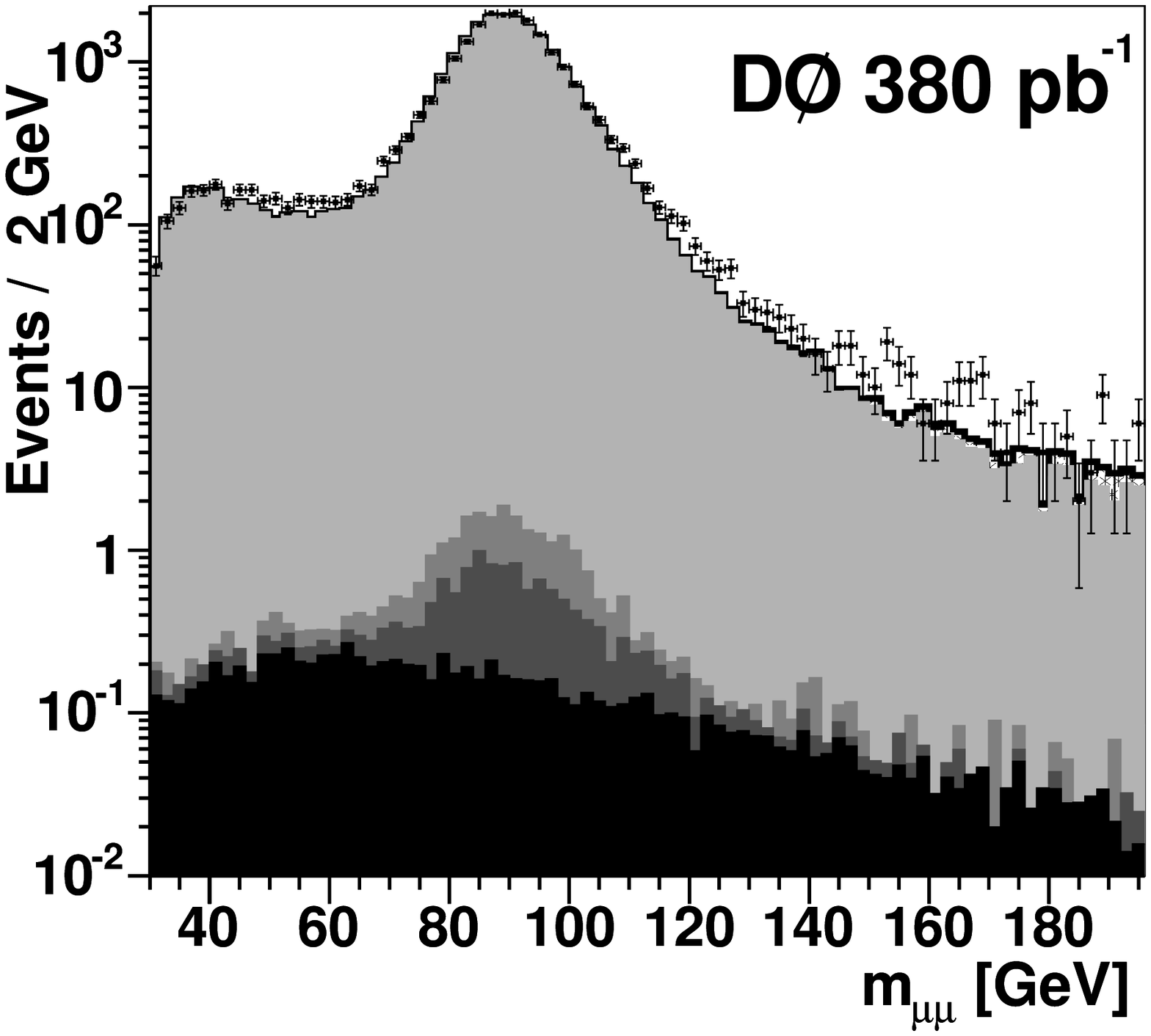}}
  \put(4.3,-0.3)  {\includegraphics[scale=0.23]{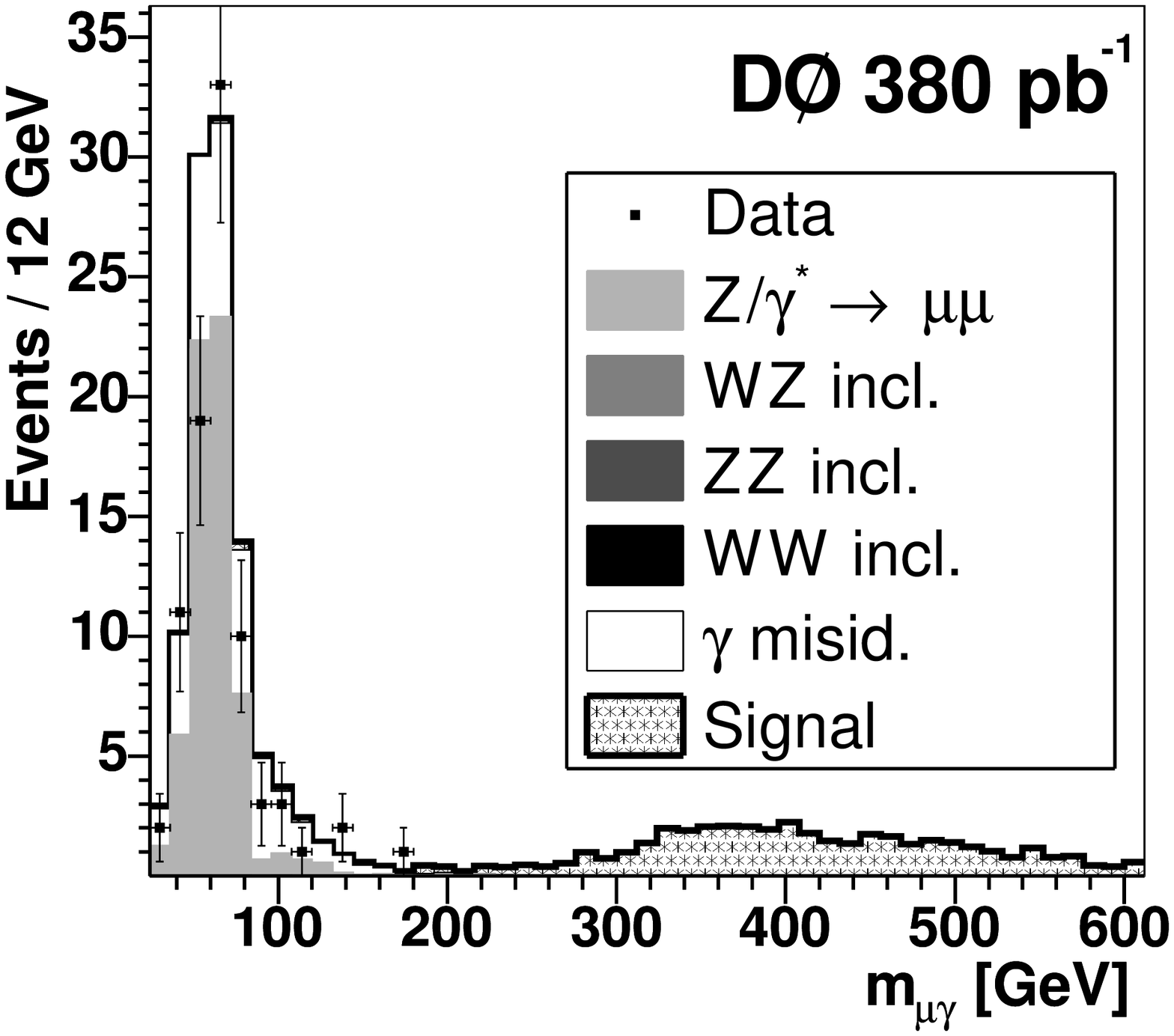}}
  \put(0.0,3.9)  {\sf a)}
  \put(4.6,3.9)  {\sf b)}
  \end{picture}
  \caption{a) Invariant dimuon mass distribution in the dimuon data sample compared to the SM
  expectation, b) invariant mass of the leading muon and the photon in the $\mu \mu \gamma$
  sample, for data (points with statistical uncertainties), SM backgrounds (DY and diboson production,
  shaded histograms, as well as the uncertainty due to jets misidentified as photons), and the expected
  signal for $m_{\mu^*} = 400$~GeV and $\Lambda = 1$~TeV.}
  \label{control1}
\end{figure}

Timing information from the muon scintillation counters is used in order to reject
cosmic ray background. Since the signal is expected to produce isolated muons,
at least one of the muons is required to be isolated: the amount of energy
deposited in the calorimeter along the muon direction in a hollow cone with
inner radius $\Delta{\cal R}=0.1$ ($\Delta{\cal R} = \sqrt{(\Delta\eta)^2 +
(\Delta\phi)^2}$) and outer radius $\Delta{\cal R}=0.4$ is required to be less
than $2.5$~GeV, and the sum of the transverse momenta of tracks within a cone of
$\Delta{\cal R}=0.5$ has to be below $2.5$~GeV, excluding the muon track. The
cumulative efficiency of the muon and track reconstruction and muon
identification is found to be $88 \pm 4$\% per muon, and the isolation condition
is $95 \pm 4$\% efficient.
The selected dimuon sample contains 24853 events, whereas 
$23200 \pm 2700$ events are expected from DY processes, and $34 \pm 4$ events are
expected from diboson production. 
The invariant dimuon mass distribution is shown in Fig.~\ref{control1} a).

Next, a photon is identified in the event as an isolated cluster of 
calorimeter energy with a characteristic shower shape and at least 90\% of the energy
deposited in the electromagnetic section of the calorimeter. The isolation condition is
$(E_{\rm tot}(0.4) - E_{\rm em}(0.2)) / E_{\rm em}(0.2) < 0.15$, where $E_{\rm tot}(0.4)$
and $E_{\rm em}(0.2)$ denote the energy deposited in the calorimeter and only its
electromagnetic section in cones of size $\Delta{\cal R}=0.4$ and 0.2, respectively. The
transverse energy $E_T$ must be larger than 16~GeV, no track is allowed to be matched to
the photon candidate with a $\chi^2$ probablility of greater than $0.1$\%, and the sum of
the transverse momenta of tracks within a hollow cone defined by $0.05 < \Delta{\cal R} <
0.4$ around the photon direction has to be below 2~GeV to further ensure
isolation. The photon candidate is required to be separated from the muon candidates in
the event by at least $\Delta{\cal R}= 0.4$, and has to be reconstructed within the
central part of the calorimeter ($|\eta|<1.1$).

\begin{figure} \setlength{\unitlength}{1cm}
  \begin{picture}(8.6,4.2)(0.0,0.0)
  \put(-0.3,-0.3) {\includegraphics[scale=0.23]{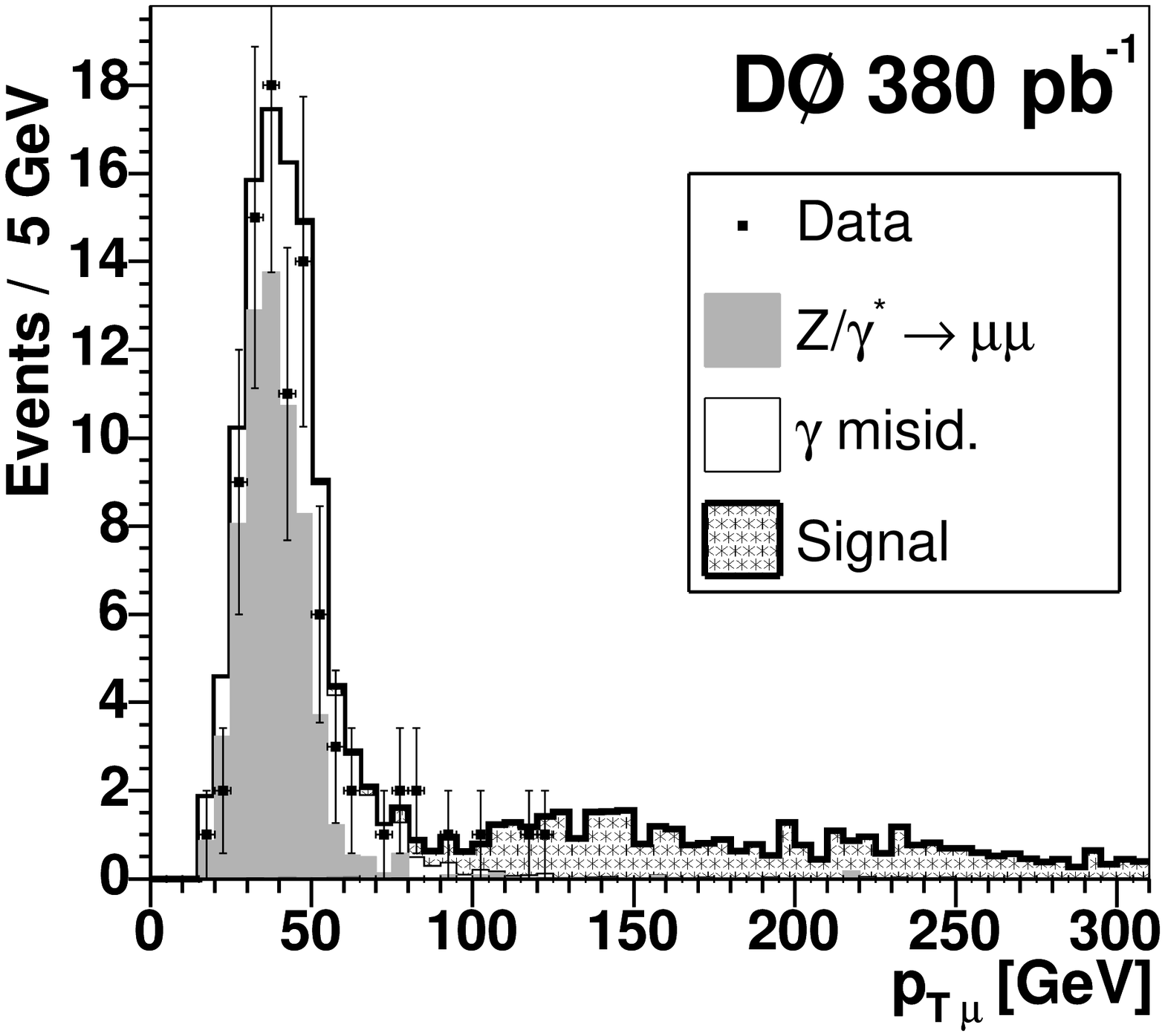}}
  \put(4.3,-0.3) {\includegraphics[scale=0.23]{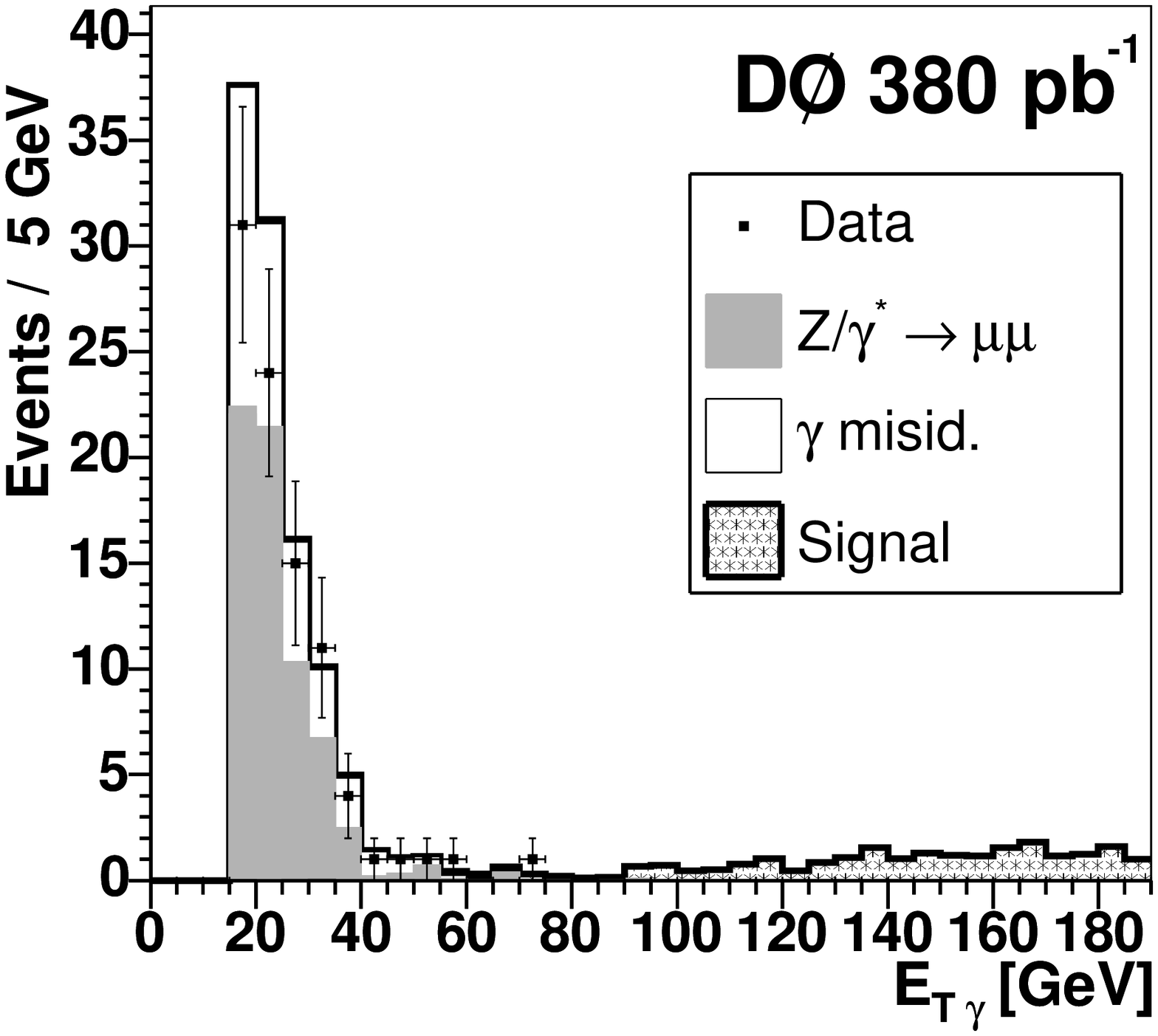}}
  \put(0.0,3.9)  {\sf a)}
  \put(4.6,3.9)  {\sf b)}
  \end{picture}
  \caption{For the $\mu \mu \gamma$ sample, a) the distribution of the leading muon
  $p_T$, and b) the photon $E_T$. Shown are the data as points with statistical uncertainties,
  the dominant SM background (DY, shaded histogram, also shown is the uncertainty due to jets
  misidentified as photons), and the expected signal for $m_{\mu^*} = 400$~GeV and
  $\Lambda = 1$~TeV.}
  \label{control2}
\end{figure}

After this selection, we expect $65 \pm 8$ events from DY processes, and less than one
event from diboson production. To estimate the possible additional background from jets
misidentified as photons and not included in the simulation, the misidentification rate has
been determined from an inclusive jet data sample; this rate applied to the dimuon plus jet
sample results in $39 \pm 5$ such events in the $\mu\mu\gamma$ selection. As a function of
$E_T$, the photon fake rate is about $0.5\%$ per jet at low $E_T$, and is negligible above
$\approx 80$~GeV. The background from jets misidentified as photons is treated as a
systematic uncertainty, resulting in a total SM expectation of $65 \pm 8 \,^{+39}_{-0}$
events. We find 90 events in the data, in good agreement with the expectation. The
invariant mass of the leading muon and the photon is shown in Fig.~\ref{control1} b) for
the data, SM expectation, and signal expectation for $m_{\mu^*} = 400$~GeV and $\Lambda =
1$~TeV. The $p_T$ distribution of the leading muon and the $E_T$ distribution of the photon
are shown in Fig.~\ref{control2}.


Additional selection criteria are applied to reduce the remaining SM background. The photon $E_T$
is required to be larger than 27~GeV. The efficiency to identify a photon is constant
at about 90\% above this value.
The final discriminant to suppress remaining SM backgrounds is the invariant mass of
the leading muon and the photon. For masses $m_{\mu^*}$ above $\approx 300$~GeV, the
leading muon is predominantly the
muon from the $\mu^*$ decay. In order to maximize the sensitivity of the analysis,
the signal expectation is calculated for $\Lambda = 1$~TeV, the background including
DY processes and diboson production is considered, and a cut value is chosen for each
value of $m_{\mu^*}$. The result is shown in Table \ref{results} along with the SM
expectation for the number of data events and the signal efficiency, which varies
between 8\% and 15\%.

\begin{table}
  \begin{center}
  \begin{tabular}{c|c|c|r@{$\,\pm \,$}l|r@{$\,\pm \,$}l}
    $m_{\mu^*}$~     & ~$m_{\mu\gamma}$ cut~ & ~Data~      & \multicolumn{2}{c|}{SM}          & \multicolumn{2}{c}{Signal eff.} \\
    \,[GeV]          & [GeV]                 &             & \multicolumn{2}{c|}{expectation} & \multicolumn{2}{c}{[\%]}        \\ \hline
    100              &                   200 &    0	   & ~$0.170$ & $0.126$~      & ~$7.5$  & $1.0$~  \\
    200              &                   200 &    0	   &  $0.170$ & $0.126$       & ~$12.5$ & $1.5$   \\
    300              &                   280 &    0	   &  $0.041$ & $0.023$       &  $12.1$ & $1.5$   \\
    400              &                   330 &    0	   &  $0.016$ & $0.011$       &  $14.7$ & $1.8$   \\
    500              &                   440 &    0	   &  $0.003$ & $0.001$       &  $11.9$ & $1.5$   \\
    600              &                   440 &    0	   &  $0.003$ & $0.001$       &  $14.4$ & $1.8$   \\
    700              &                   440 &    0	   &  $0.003$ & $0.001$       &  $13.6$ & $1.7$   \\
    800              &                   440 &    0	   &  $0.003$ & $0.001$       &  $14.5$ & $1.8$   \\
    900              &                   440 &    0	   &  $0.003$ & $0.001$       &  $14.7$ & $1.8$   \\
    1000             &                   440 &    0	   &  $0.003$ & $0.001$       &  $14.4$ & $1.8$   \\
  \end{tabular}
  \caption{For different values of $m_{\mu^*}$, the final selection requirement on the invariant mass of the
           leading muon and the photon, the remaining data events, the SM expectation, and the signal
	   efficiency. The quoted uncertainties include statistical and systematic uncertainties added
	   in quadrature.}
  \label{results}
  \end{center}
\end{table}

The dominant systematic uncertainties are as follows. The uncertainty on the SM cross
sections is dominated by the DY process and the uncertainty from the choice of PDF and
renormalization and factorization scales (4\%). Muon reconstruction and identification
have an uncertainty of 4\% per muon, and a 3\% error is assigned to the photon
identification. The uncertainty due to the trigger efficiency is 7\%. The integrated
luminosity is known to a precision of 6.5\% \cite{d0lumi}. The uncertainty due to jets
misidentified as photons is dominant after all selection criteria for $m_{\mu^*}$ up to $400$~GeV:
for $m_{\mu^*} = 100$~GeV ($400$~GeV), 0.097 (0.008) such ``fake" photons are
expected, while for $m_{\mu^*} = 500$~GeV and above this background is negligible ($<
10^{-5}$ events). The uncertainty on the signal cross section is estimated to be 10\%,
consisting of PDF uncertainties and unknown higher order corrections.

\begin{figure} \setlength{\unitlength}{1cm}
  \begin{picture}(8.6,5.2)(0.0,0.0)
  \put(1.5,-0.3) {\includegraphics[scale=0.30]{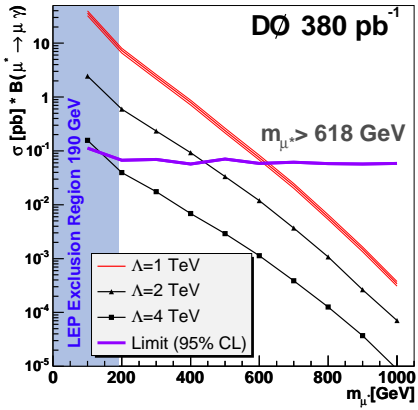}}
  \end{picture}
  \caption{The measured cross section $\times$ branching fraction limit, compared to
  the contact interaction model prediction for different choices of $\Lambda$. For the
  case $\Lambda = 1$~TeV, the theoretical uncertainty of the model prediction is indicated.}
  \label{limit1}
\end{figure}

\begin{figure} \setlength{\unitlength}{1cm}
  \begin{picture}(8.6,5.3)(0.0,0.0)
  \put(1.5,-0.3) {\includegraphics[scale=0.30]{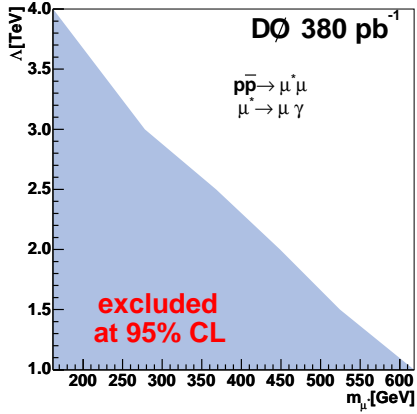}}
  \end{picture}
  \caption{The region in the plane of $\Lambda$ and $m_{\mu^*}$ excluded by the
  present analysis.}
  \label{limit2}
\end{figure}

Since no events are found in the data, in agreement with the SM expectation, we
set 95\% confidence level limits on the $\mu^*$ production cross section times
the branching fraction into $\mu \gamma$. A Bayesian technique \cite{d0limit} is used,
taking into account all uncertainties and treating them as symmetric for simplicity.
The resulting limit as a function of $m_{\mu^*}$
is shown in Fig.~\ref{limit1} together with predictions of the contact interaction
model for different choices of the scale $\Lambda$. For $\Lambda = 1$~TeV
($\Lambda = m_{\mu^*}$), masses below 618~GeV (688~GeV) are excluded.
In Fig.~\ref{limit2} the excluded region in terms of $\Lambda$ and $m_{\mu^*}$ is shown.

The CDF collaboration has recently searched \cite{cdf} for the production of excited
electrons, and obtained comparable cross section limits, but the CDF mass limit of
$m_{e^*} > 879$~GeV at 95\% C.L.~for $\Lambda = m_{e^*}$ cannot be directly compared to
ours for two reasons. The cross section calculated with the version of {\sc pythia} used
by CDF is a factor of two higher than in subsequent versions corrected by the {\sc
pythia} authors. Furthermore, CDF assumes that decays via contact interactions can be
neglected, while in our analysis such decays are taken into account in the calculation
of the branching fraction $\mu^* \rightarrow \mu \gamma$, following \cite{Baur90,LHC}.
If we adjusted our result for these two differences, we would obtain a limit of
$m_{\mu^*} > 890$~GeV at 95\% C.L.~for $\Lambda = m_{\mu^*}$.

In summary, we have searched for the production of excited muons in the process
$p \bar p \rightarrow \mu^* \mu$ with $\mu^* \rightarrow \mu \gamma$, using
$380$~pb$^{-1}$ of data collected with the D0 detector. We find no
events in the data, compatible with the SM expectation, and set limits on
the production cross section times branching fraction as a function of the mass of the
excited muon. For a scale parameter $\Lambda = 1$~TeV, masses below 618~GeV are
excluded, representing the most stringent limit to date.

We thank A. Daleo and M. Kr\"amer for useful discussions, and A. Daleo for providing
us with the NNLO corrections to the $\mu^*$ production cross section.
%
We thank the staffs at Fermilab and collaborating institutions, 
and acknowledge support from the 
DOE and NSF (USA);
CEA and CNRS/IN2P3 (France);
FASI, Rosatom and RFBR (Russia);
CAPES, CNPq, FAPERJ, FAPESP and FUNDUNESP (Brazil);
DAE and DST (India);
Colciencias (Colombia);
CONACyT (Mexico);
KRF and KOSEF (Korea);
CONICET and UBACyT (Argentina);
FOM (The Netherlands);
PPARC (United Kingdom);
MSMT (Czech Republic);
CRC Program, CFI, NSERC and WestGrid Project (Canada);
BMBF and DFG (Germany);
SFI (Ireland);
The Swedish Research Council (Sweden);
Research Corporation;
Alexander von Humboldt Foundation;
and the Marie Curie Program.
%


\begin{thebibliography}{99}
\bibitem{coordinate} The pseudorapidity $\eta$ is defined as $\eta = - \ln (\tan(\theta / 2))$.
            We use the polar angle $\theta$ relative to the proton beam direction, and $\phi$
            is the azimuthal angle, all measured with respect to the geometric center of
            the detector.
\end{thebibliography}
\end{document}